\documentclass[10pt,twocolumn,oneside]{IEEEtran} 

\usepackage{cite}

\usepackage{amssymb}
\usepackage{amsmath}

\usepackage{theorem}
\usepackage{graphicx}
\graphicspath{{./graphics/}{.}}
\usepackage{color}

\usepackage{multirow}

\newcommand{\Complex}{\mathbb{C}}

\long\def\symbolfootnote[#1]#2{\begingroup%}
\def\thefootnote{\fnsymbol{footnote}}\footnote[#1]{#2}\endgroup}
\newcommand{\Expected}{\mathbb{E}}

\newcommand{\bsm}[1]{{\boldsymbol #1}}
\newcommand{\hd}{\hdots}

\newtheorem{thm}{Theorem}

\newtheorem{lem}{Lemma}

\IEEEoverridecommandlockouts

\begin{document}
\title {Partially-blind Estimation of Reciprocal Channels for AF Two-Way Relay Networks Employing \textit{M}-PSK Modulation}
\author{Saeed Abdallah and Ioannis N. Psaromiligkos\thanks{The authors are with the Electrical and Computer Engineering Dept., McGill University, Montreal, Quebec, Canada. Email: saeed.abdallah@mail.mcgill.ca; yannis@ece.mcgill.ca.}\\
\thanks{This work was supported in part by the Natural Sciences and Engineering Research Council (NSERC) under grant 262017.}}

\maketitle

\begin{abstract}

We consider the problem of channel estimation for amplify-and-forward two-way relays assuming channel reciprocity and \textit{M}-PSK modulation. In an earlier work, a partially-blind maximum-likelihood estimator was derived by treating the data as deterministic unknowns. We prove that this estimator approaches the true channel with high probability at high signal-to-noise ratio (SNR) but is not consistent. We then propose an alternative estimator which is consistent and has similarly favorable high SNR performance. We also derive the Cramer-Rao bound on the variance of unbiased estimators. 

\begin{center}\textbf{\textit{Index Terms}:} Amplify-and-forward, Channel estimation, Cramer-Rao bound, Two-way relay networks. \end{center}

\end{abstract}

\section{Introduction}

Amplify-and-forward (AF) Two-Way Relay Networks (TWRNs)~\cite{rankov07} have received a lot of attention for their ability to improve the spectral efficiency of bidirectional communication. An essential requirement in these networks is reliable channel knowledge at both ends of the communication link. The channel estimation problem for AF two-way relays has been considered in several recent works~\cite{gao2009optimal,gao2009channel,tensor_based}. These works developed training-based algorithms which estimate the channel using pilot symbols known to both terminals. Unfortunately, the required training overhead may significantly reduce the spectral efficiency of the system. It is therefore worthwhile to develop reliable partially or completely blind channel estimation approaches that minimize this loss of spectral efficiency. 

In~\cite{ciss_paper} and~\cite{msev_arxiv}, the authors developed partially-blind maximum-likelihood (ML)-based channel estimation algorithms for TWRNs assuming that the channels are flat fading and that the two terminals employ $M$-PSK modulation. In both works, the data symbols were treated as deterministic unknowns and pilots were only needed to resolve the phase ambiguity. The difference between~\cite{ciss_paper} and~\cite{msev_arxiv} is that the channels were assumed to be reciprocal in~\cite{ciss_paper} but nonreciprocal in~\cite{msev_arxiv}. It was proved in~\cite{msev_arxiv} that the ML estimator for nonreciprocal channels is consistent and that it approaches the true channel with high probability at high SNR for $M>2$. 

In this work, we focus on reciprocal channels. We analyze the asymptotic high SNR behavior of the ML estimator developed in~\cite{ciss_paper} and investigate its consistency. We show that for $M>2$, the ML estimator approaches the true channel with high probability at high SNR. However, we also prove that it is not consistent. No such analysis was provided in~\cite{ciss_paper}. 

As an alternative to the ML estimator, we propose to estimate the channel by minimizing the sample variance of the envelope of the received signal after self-interference cancellation. This criterion was used by the ML estimator developed for nonreciprocal channels in~\cite{msev_arxiv}. In this work, we investigate its application to reciprocal channels. We term the proposed estimator as the minimum sample envelope variance (MSEV) estimator. The asymptotic behavior of the estimator established in~\cite{msev_arxiv} still holds under channel reciprocity. Thus, the MSEV estimator is consistent and approaches the true channel with high probability at high SNR for $M>2$. 

As a reference, we also derive two Cramer-Rao bounds (CRBs) on the variance of unbiased estimators under channel reciprocity. Monte-Carlo simulations are then used to obtain the mean-squared error (MSE) of the two estimators, demonstrating that the proposed estimator outperforms the ML estimator. In summary, the main contributions of this work are: (\textit{i}) analysis of the high SNR performance of the ML estimator for reciprocal channels; (\textit{ii}) investigation of the consistency of the ML estimator for reciprocal channels; (\textit{iii}) application of the MSEV criterion to reciprocal channels; (\textit{iv}) derivation of two CRBs on the variance of unbiased estimators for reciprocal channels.

The remainder of this correspondence is organized as follows. In Section~\ref{system_model}, we present our system model. In Section~\ref{proposed_algorithm}, we present the ML estimator of~\cite{ciss_paper} and the proposed estimator. In Section~\ref{performance_analysis}, we analyze the high SNR behavior and the consistency of the two estimators. The CRBs are derived in Section~\ref{CRB}. Our simulation results are shown in Section~\ref{simulations}. Finally, our conclusions are in Section~\ref{conclusions}.
\vspace{0.5ex}
\begin{figure}[htbp]
\centering
\includegraphics[height=1.0in,width=3.2in]{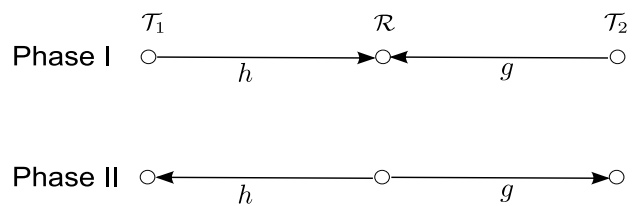}
\caption{The two-way relay network with two source terminals and one relay, assuming channel reciprocity.}
\label{TWRN_reciprocal}
\end{figure}
\vspace{0.5ex}

\section{System Model}
\label{system_model}

We consider the half-duplex TWRN with two source terminals, $\mathcal{T}_1$ and $\mathcal{T}_2$, and a single relay, $\mathcal{R}$, shown in Fig.~\ref{TWRN_reciprocal}. The network operates in quasi-static flat-fading channel conditions. Each data transmission period is divided into two phases~\cite{rankov07}. In the first phase, $\mathcal{T}_1$ and $\mathcal{T}_2$ simultaneously transmit to $\mathcal{R}$ the $M$-PSK data symbols $t_{1}=\sqrt{P_1}e^{\jmath\phi_{1}}$ and $t_{2}=\sqrt{P_2}e^{\jmath\phi_{2}}$, respectively, where $P_1$ and $P_2$ are the transmission powers, and $\phi_{1}$ and $\phi_{2}$ are chosen from the set $S_M=\{(2\ell-1)\pi/M,\ell=1,\hdots,M\}$. The received signal at the relay during this phase is given by $r=ht_{1}+gt_{2}+n$ where $h$ and $g$ are the complex coefficients representing the flat-fading channels $\mathcal{T}_1\rightarrow\mathcal{R}$ and $\mathcal{T}_2\rightarrow\mathcal{R}$, respectively, and $n$ is the zero-mean complex additive white Gaussian noise (AWGN) with variance $\sigma^2$. In the second phase, the relay broadcasts the amplified signal $Ar$, where\footnote{This choice of the amplification factor maintains an average power of $P_r$ at the relay over the long term~\cite{gao2009optimal}.} $A=\sqrt{\frac{P_r}{P_1+P_2+\sigma^2}}$. Without loss of generality, we will consider estimation at $\mathcal{T}_1$. Assuming channel reciprocity, the received signal at $\mathcal{T}_1$ during the second phase is
\begin{equation}
\label{atT1}
z=Aat_{1}+Abt_{2}+Ahn+\eta,
\end{equation}
where $a\triangleq h^2$, $b\triangleq gh$, and $\eta$ is the zero-mean complex AWGN term at $\mathcal{T}_1$ with variance $\sigma^2$. Since, $M$-PSK modulation is assumed, it is sufficient for detection to know $a$ and $\phi_b\triangleq\angle b$.

\section{Channel Estimation Algorithms}
\label{proposed_algorithm}

Let the vectors $\bsm{z}\triangleq[z_1\hd z_N]^{T}$, $\bsm{t}_1\triangleq[t_{11},\hd,t_{1N}]^{T}$, $\bsm{t}_2\triangleq[t_{21},\hd,t_{2N}]^{T}$, $\bsm{n}=[n_1,\hdots n_N]$, and $\bsm{\eta}=[\eta_1,\hd,\eta_N]$ denote the received vector at $\mathcal{T}_1$, the transmitted symbol vectors of $\mathcal{T}_1$ and $\mathcal{T}_2$, the noise vector at $\mathcal{R}$ and the noise vector at $\mathcal{T}_1$ during $N$ successive transmissions, respectively. We note that the sample size $N$ is constrained by the coherence time of the channel during which the channel parameters $a$ and $b$ remain constant.

\subsection{Maximum Likelihood Estimation}
We begin by briefly presenting the ML estimator for reciprocal channels derived in~\cite{ciss_paper}. To avoid dealing with a complicated likelihood function, the transmitted symbols $t_{2i}, i=1,\hdots,N$ are treated as deterministic unknowns. We also ignore the finite alphabet constraint that restricts the phases $\phi_{2i}\triangleq\angle t_{2i}$, $i=1\hd, N$ to the set $S_{M}$. The ML approach can be used to blindly estimate the parameter $a$ and the sums $\psi_{i}\triangleq\phi_{2i}+\phi_b,\ i=1,\hdots,N$. However, it cannot be used to obtain separate estimates of $\phi_b$ and $\phi_{2i}$, $i=1,\hd, N$. Hence, a small number of pilots will be needed to isolate an estimate of $\phi_b$, and because of this we will refer to the ML estimator as partially blind. 
The unknown parameters are collected in the vector $\boldsymbol{\theta}\triangleq[a, |b|, \psi_1,\hdots,\psi_{N}]^{T}$. Let $\hat{a}$, $\widehat{|b|}$, $\hat{\psi}_{i}$ be the ML estimates of $a$, $|b|$, and $\psi_{i}$, $i=1,\hdots,N,$ respectively. It is shown in~\cite{ciss_paper} that $\hat{\psi}_{i}=\angle(z_{i}-A\hat{a}t_{1i})$, $\widehat{|b|}=\frac{1}{NA\sqrt{P_2}}\sum\limits_{i=1}^{N}\left|z_i-A\hat{a}t_{1i}\right|$, and that $\hat{a}$ is given by
\begin{equation}
\label{a_ML}
\begin{split}
\hat{a}=&\arg\!\min_{u\in\Complex}\Bigg\{\frac{\sum\limits_{i=1}^{N}\left(|z_i-Aut_{1i}|-\frac{1}{N}\sum\limits_{k=1}^{N}|z_k-Aut_{1k}|\right)^2}{\sigma^2(A^2|u|+1)}\\
&\ \ \ \ \ \ \ \ \ \ \ \ \ \ \ \ \ \ \ \ \ +N\log\left(A^2|u|+1\right)\Bigg\}.
\end{split}
\end{equation}
We show in Section~\ref{performance_analysis} that the estimator in~\eqref{a_ML} is not consistent. We next propose an alternative criterion which results in a consistent estimator of $a$. 

\subsection{Proposed MSEV Algorithm}

Let $\tilde{z}_i(u)\triangleq \left(z_i-Aut_{1i}\right),\ i=1,\hdots,N$ be the ``cleaned''  versions of the received signal samples after self-interference has been removed using the complex value $u$ as an estimate of $a$. The signals $\tilde{z}_i(u),\ i=1,\hdots,N$ are independently generated realizations of the random variable $\tilde{z}(u)\triangleq A(a-u)t_1+Abt_2+Ahn+\eta$. The quantity\footnote{$W_N(u)$ is a scaled version of the numerator of the first term in~\eqref{a_ML}.}
\begin{equation}
\label{sample_variance}
W_N(u)\triangleq\frac{1}{N-1}\sum\limits_{i=1}^{N}\left(|\tilde{z}_i(u)|-\frac{1}{N}\sum\limits_{k=1}^{N}|\tilde{z}_k(u)|\right)^2
\end{equation}
represents the sample variance of the envelope of $\tilde{z}(u)$. One would intuitively expect the actual variance of $|\tilde{z}(u)|$ to be smallest in the absence of self-interference (i.e., at $u=a$). We therefore propose the MSEV estimator of $a$ which is given by
\begin{equation}
\label{a_MSEV}
\hat{a}_{v}=\arg\!\min_{u\in\Complex}\hspace{0.8ex}\frac{1}{N-1}\sum\limits_{i=1}^{N}\left(|\tilde{z}_i(u)|-\frac{1}{N}\sum\limits_{k=1}^{N}|\tilde{z}_k(u)|\right)^2.
\end{equation}
The MSEV estimator has the same form as the ML estimator of the self-interference channel for the nonreciprocal case in~\cite{msev_arxiv}. However, it does not retain the same maximum-likelihood interpretation when channel reciprocity is assumed. 

The solutions for~\eqref{a_ML} and~\eqref{a_MSEV} may be obtained using iterative methods such as the steepest-descent algorithm or the quasi-Newton algorithm~\cite{boyd2004convex}. Since the objective functions in~\eqref{a_ML} and~\eqref{a_MSEV} are nonconvex, the performance of such methods will depend on the availability of good initial estimates. A simple way to obtain an initial estimate is the sample average estimator used in~\cite{msev_arxiv} and given by $\hat{a}_s=\frac{1}{NAP_1}\sum_{i=1}^{N}t_{1i}^{*}z_i$. In our experimental studies, we will use the steepest descent algorithm to solve~\eqref{a_ML} and~\eqref{a_MSEV}. As we shall see, the resulting MSE for both estimators is almost identical to that obtained when the solutions are acquired using grid search. In the next section, we compare the asymptotic behavior of the two estimators. 

\section{Asymptotic Behavior Analysis}
\label{performance_analysis}

Our analysis focuses on the estimation of $a$ since $|b|$ is not required for detection. We study the asymptotic behavior of the two estimators by considering their consistency and their behavior at high transmit SNR. The transmit SNR is defined as $\gamma\triangleq\frac{P_2}{\sigma^2}$. As we shall see, both $\hat{a}$ and $\hat{a}_v$ approach $a$ as $\gamma\rightarrow\infty$ with high probability for $M\geq3$. Moreover, the MSEV estimator is consistent while the ML estimator is not. We start our analysis with the MSEV estimator.

\subsection{High SNR Behavior of the MSEV Estimator}

It can be easily verified that the high SNR behavior of the ML estimator under nonreciprocal channels in~\cite{msev_arxiv} (see Lemma 1 and Lemma 2 therein) is not affected when channel reciprocity is assumed. Hence, letting $G_{\gamma}(u)=\frac{1}{\sigma^2}W_N(u)$, and assuming that $P_1=\alpha P_2$ and $P_r=\beta(P_1+P_2)$ for some $\alpha,\beta>0$, the following theorem holds.

\begin{thm}
\label{theorem1}
For fixed $N$, and assuming $M$-PSK modulation, $\lim\limits_{\gamma\rightarrow\infty}G_{\gamma}(u)=+\infty$ for all $u\neq a$ with probability $1-\left(\frac{2}{M}\right)^{N-1}(M-1)$. For $u=a$, $\lim\limits_{\gamma\rightarrow\infty}G_{\gamma}(a)<+\infty$.
\end{thm}
The above theorem guarantees that for $M\geq3$ and for sufficiently large $N$, the MSEV estimator will approach the true channel with high probability as SNR increases.

\subsection{High SNR Behavior of the ML Estimator}

Denoting by $\Lambda_{\gamma}(u)$ the objective function of the ML estimator in~\eqref{a_ML}, we have that
\begin{equation}
\begin{split}
\Lambda_{\gamma}(u)&=NG_{\gamma}(u)/(A^2|u|+1)+N\log\left(A^2|u|+1\right)\\
&=\frac{N\left((1+\alpha)\gamma+1\right)G_{\gamma}(u)}{\beta(1+\alpha)\gamma|u|+(1+\alpha)\gamma+1}\\
&\ \ +N\log\left[\frac{\beta(1+\alpha)\gamma|u|}{(1+\alpha)\gamma+1}+1\right].
\end{split}
\end{equation}
Hence,
\begin{equation}
\begin{split}
\lim\limits_{\gamma\rightarrow\infty}\Lambda_{\gamma}(u)=&\frac{N(1+\alpha)}{\beta(1+\alpha)|u|+(1+\alpha)}\lim\limits_{\gamma\rightarrow\infty}G_{\gamma}(u)\\
&\ +N\log\left(\beta|u|+1\right).
\end{split}
\end{equation}
Thus, by Theorem~\ref{theorem1}, $\lim\limits_{\gamma\rightarrow\infty}\Lambda_{\gamma}(a)<+\infty$ and $\lim\limits_{\gamma\rightarrow\infty}\Lambda_{\gamma}(u)=+\infty$ for all $u\neq a$ with probability $1-\left(\frac{2}{M}\right)^{N-1}(M-1)$, i.e., the ML estimator exhibits similar high SNR behavior to the MSEV estimator. 

\subsection{Consistency of the MSEV Estimator}

The proof for the consistency of the ML estimator under nonreciprocal channels in~\cite{msev_arxiv} is not affected when channel reciprocity is assumed. Since $W_N(u)$ is the sample variance of $|\tilde{z}(u)|$, it converges in probability to the variance of $|\tilde{z}(u)|$, which is given by~\cite{msev_arxiv}:
\begin{equation}  
\label{var_exp_ciss} 
\begin{split}
\mathcal{W}(u)&=A^2|a-u|^2P_1+A^2|b|^2P_2+\sigma^2(A^2|a|+1)\\
&-\frac{\pi\sigma^2(A^2|a|+1)}{4M^2}\left(\sum_{k=0}^{M-1}L_{\text{\tiny$1/2$}}\left(-\lambda_{k}(a-u)\right)\right)^2
\end{split}
\end{equation}
where $L_{\text{\tiny$1/2$}}(x)$ is the Laguerre polynomial~\cite{abramowitz} with parameter $1/2$,
$I_{\tau}(\cdot)$ is the Modified Bessel Function of the First Kind of order $\tau$, and 
\begin{equation}
\label{lambda_k_v}
\begin{split}
\lambda_{k}(v)&\triangleq\frac{1}{\sigma^2(A^2|a|^2+1)}\big(A^2|v|^2P_1+A^2|b|^2P_2\\
&+2A^2|v||b|\sqrt{P_1P_2}\cos\left(\angle v-\phi_b+2\pi k/M\right)\big).
\end{split}
\end{equation}
The variance $\mathcal{W}(u)$ has a unique global minimum at $u=a$, and the estimator is consistent when the channel parameters $a, b$ belong to compact sets. 

\subsection{Consistency of the ML Estimator}
Let $Y_N(u)$ be the objective function of the ML estimator in~\eqref{a_ML} scaled by $\frac{1}{N-1}$. Thus, 
\begin{equation}
\label{con_ciss1}
Y_N(u)\triangleq\frac{W_N(u)}{\sigma^2(A^2|u|+1)}+\frac{N}{N-1}\log\left(A^2|u|+1\right).
\end{equation}
As $N\rightarrow\infty$, $Y_N(u)$ converges in probability to
\begin{equation}
\mathcal{Y}(u)\triangleq\frac{\mathcal{W}(u)}{\sigma^2(A^2|u|+1)}+\log\left(A^2|u|+1\right).
\end{equation} 
To find out whether $\mathcal{Y}(u)$ has an extremum at $u=a$, we analyze the behavior of the partial derivatives\footnote{In this work, we use $\Re\{\cdot\}$ and $\Im\{\cdot\}$ to denote the real and imaginary parts of a complex number, respectively.} $\dot{\mathcal{Y}}_{\Re}(u)\triangleq\frac{\partial\mathcal{Y}(u)}{\partial\Re\{u\}}$ and $\dot{\mathcal{Y}}_{\Im}(u)\triangleq\frac{\partial\mathcal{Y}(u)}{\partial\Im\{u\}}$. Because of the symmetry of $\mathcal{Y}(u)$ with respect to $\Re\{u\}$ and $\Im\{u\}$, it is sufficient to consider $\dot{\mathcal{Y}}_{\Re}(u)$. We have
\begin{equation}
\label{ciss_derivative}
\begin{split}
\dot{\mathcal{Y}}_{\Re}(u)=&\frac{A^2\Re\{u\}}{|u|(A^2|u|+1)}\left[1-\frac{\mathcal{W}(u)}{\sigma^2(A^2|u|+1)}\right]\\
&+\frac{1}{\sigma^2(A^2|u|+1)}\frac{\partial\mathcal{W}(u)}{\partial\Re\{u\}}.
\end{split}
\end{equation}
Moreover, it can be shown that $\frac{\partial\mathcal{W}(u)}{\partial\Re\{u\}}\Big|_{u=a}\hspace{-0.1in}=0$. Therefore, 
\begin{equation}
\label{ciss_derivative_a}
\dot{\mathcal{Y}}_{\Re}(a)=\frac{A^2\Re\{a\}}{|a|(A^2|a|+1)}\left[1-\frac{\mathcal{W}(a)}{\sigma^2(A^2|a|+1)}\right].
\end{equation}
For the factor $\left[1-\frac{\mathcal{W}(a)}{\sigma^2(A^2|a|+1)}\right]$ we have the following lemma (proved in Appendix~\ref{appendix_ciss}).
\begin{lem}
\label{ciss_lem}
For any $a\in\Complex$, $\left[1-\frac{\mathcal{W}(a)}{\sigma^2(A^2|a|+1)}\right]>0$.
\end{lem}
Hence, $\dot{\mathcal{Y}}_{\Re}(a)$ is zero only when $\Re\{a\}=0$. Similarly, $\dot{\mathcal{Y}}_{\Im}(a)$ is zero only when $\Im\{a\}$ is zero. Since $\mathcal{Y}(u)$ is differentiable at $u=a$ (for $a\neq0$) and $u=a$ is not a boundary point, this implies that $\mathcal{Y}(u)$ does not have an extremum at $u=a$ for $a\neq0$. Hence, the ML estimator is not consistent~\cite{handbook_econometrics}.

The inconsistency of the ML estimator should not come as a surprise since the data symbols are treated as deterministic unknowns. Due to this assumption, the received samples are not identically distributed and the number of estimated parameters is not fixed but grows linearly with the number of samples. Hence, the well-known sufficient conditions for the consistency of ML estimators (see~\cite[Theorem 10.1.6]{statistical_inference}) are not satisfied. The unknown data symbols are referred to in the literature as incidental parameters, while the channel parameters which are the same for all samples are called structural parameters~\cite{neyman1948,mak1982}. It is well known that the asymptotic properties of ML estimators, such as consistency, do not necessarily hold in the presence of incidental parameters~\cite{neyman1948}. In fact, the ML estimator of structural parameters can be inconsistent even when a consistent estimator exists~\cite{mak1982}. 

\section{Cramer-Rao Bounds}
\label{CRB}

We derive two Cramer-Rao bounds for the estimation problem under consideration. The first bound is derived by treating the phases $\psi_1,\hd,\psi_N$ as deterministic unknowns. The vector of unknown real parameters is $\bsm{\theta}_R\triangleq[\Re\{a\},\Im\{a\}, |b|,\psi_1,\hd,\psi_N]^{T}$ and the PDF of $\bsm{z}$ is
\begin{equation}
\begin{split}
f(\bsm{z}\ |\ \bsm{\theta}_R)&=\frac{1}{\pi\sigma^2(A^2|a|+1)^{N}}\times\\ 
&\hspace{-2ex}\mbox{exp}\left(-\frac{\sum_{i=1}^{N}|z_i-Aat_{1i}-A|b|\sqrt{P_2}e^{\jmath\psi_i}|^2}{\sigma^2(A^2|a|+1)}\right).
\end{split}
\end{equation}
Let $\bsm{I}(\bsm{\theta}_R)$ be the corresponding Fisher information matrix (FIM). We have
\begin{equation}
\begin{split}
\hspace{-1ex}\bsm{I}(\bsm{\theta}_R)&=\Expected\left\{\hspace{-1ex}\left(\frac{\partial\log f(\bsm{z}\ |\ \bsm{\theta}_R)}{\partial\bsm{\theta}_R}\right)\hspace{-1ex}\left(\frac{\partial\log f(\bsm{z}\ |\ \bsm{\theta}_R)}{\partial\bsm{\theta}_R}\right)^{T}\right\}\\
&=\begin{bmatrix}
\ \bsm{A}&\bsm{B}\\
\ \bsm{B}^{T}&\bsm{C}
\end{bmatrix},
\end{split}
\end{equation}
where 
\begin{equation}
\bsm{A}=\begin{bmatrix}\frac{2A^2NP_1}{\sigma^2\left(A^2|a|+1\right)}\hspace{-0.5ex}+\hspace{-0.5ex}\frac{A^4\Re\{a\}^2N}{|a|^2(A^2|a|+1)^2}&\hspace{-2ex}\frac{A^4\Re\{a\}\Im\{a\}N}{|a|^2(A^2|a|+1)^2}\\
\frac{A^4\Re\{a\}\Im\{a\}N}{|a|^2(A^2|a|+1)^2}&\hspace{-2ex}\frac{2A^2NP_1}{\sigma^2\left(A^2|a|+1\right)}\hspace{-0.5ex}+\hspace{-0.5ex}\frac{A^4\Im\{a\}^2N}{|a|^2(A^2|a|+1)^2} 
\end{bmatrix}
\end{equation}
\begin{equation}
\hspace{-0.7ex}\bsm{B}^{T}=\frac{2A^2}{\sigma^2\left(A^2|a|+1\right)}\begin{bmatrix}
\Re\{e^{\jmath\phi_b}\bsm{t}_1^{H}\bsm{t}_2\}&\hspace{-1ex}\Im\{e^{\jmath\phi_b}\bsm{t}_1^{H}\bsm{t}_2\}\\
\Im\{b^*t_{11}t_{21}^{*}\}&\hspace{-1ex}\Re\{b^*t_{11}t_{21}^{*}\}\\
\vdots&\ \vdots\\
\Im\{b^*t_{1N}t_{2N}^{*}\}&\hspace{-1ex}\Re\{b^*t_{1N}t_{2N}^{*}\}
\end{bmatrix},
\end{equation}
and 
\begin{equation}
\bsm{C}=\frac{1}{\sigma^2\left(A^2|a|+1\right)}\left[2A^2NP_2, 2A^2|b|^2P_2,\hd, 2A^2|b|^2P_2\right]^{T}\hspace{-1ex}.
\end{equation}
Assuming $\bsm{I}(\bsm{\theta}_R)$ is invertible, the CRB for the estimation of $a$ is given by the sum of the first two diagonal entries in the inverse of $\bsm{I}(\bsm{\theta}_R)$, i.e.\footnote{The notation $[\bsm{A}]_{ij}$ is used to refer to the $(i,j)$th element of the matrix $\bsm{A}$.}, $CRB_a=[\bsm{I}^{-1}(\bsm{\theta}_R)]_{11}+[\bsm{I}^{-1}(\bsm{\theta}_R)]_{22}$. Let $\bsm{D}$ be the $2\times2$ top left submatrix of $\bsm{I}^{-1}(\bsm{\theta}_R)$. Using the Schur-complement property, we have that $\bsm{D}=\left(\bsm{A}-\bsm{B}\bsm{C}^{-1}\bsm{B}^{T}\right)^{-1}$, i.e.,
\begin{equation}
\label{schur}
CRB_a=\mbox{tr}\left(\left(\bsm{A}-\bsm{B}\bsm{C}^{-1}\bsm{B}^{T}\right)^{-1}\right).
\end{equation}
Because the symbols $\bsm{t}_1$ are known and the symbols $\bsm{t}_2$ are treated as deterministic unknowns, $CRB_a$ is a function of $\bsm{t}_1$ and $\bsm{t}_2$, and it thus applies for the particular realizations of $\bsm{t}_1$ and $\bsm{t}_2$ under consideration. 

Another variant of the CRB commonly used in the presence of random nuisance parameters is the Modified CRB (MCRB)~\cite{gini2000}. In deriving the MCRB, we compute a modified FIM (MFIM) by first extracting the submatrix of the conventional FIM which corresponds only to the nonrandom parameters ($\Re\{a\}$, $\Im\{a\}$ and $|b|$) and then obtaining the expectation of this submatrix with respect to the random nuisance parameters ($\bsm{t}_2$ in our case). Let $\bsm{\theta}'\triangleq[\Re\{a\}, \Im\{a\}, |b|]^{T}$, and let $\bsm{I}(\bsm{\theta}')$ be the corresponding $3\times3$ submatrix of $\bsm{I}(\bsm{\theta}_R)$. Then the MFIM is given by (19). The MCRB on $a$ is the sum of the first two diagonal entries in the inverse of $\bsm{I}_m(\bsm{\theta}')$, i.e.,  
\setcounter{equation}{19}
\begin{equation}
\begin{split}
MCRB_a&=[\bsm{I}_m^{-1}(\bsm{\theta}')]_{11}+[\bsm{I}_m^{-1}(\bsm{\theta}')]_{22}\\
&=\frac{4\sigma^2P_1(A^2|a|+1)^2+\sigma^4A^2(A^2|a|+1)}{4NA^2P_1^2(A^2|a|+1)+2N\sigma^2A^4P_1}.
\end{split}
\end{equation}
While the above bound has the advantage of a simple closed form, it is not as tight as $CRBa$.
\begin{figure*}[!t]

\normalsize

\setcounter{equation}{18}

\begin{equation}
\label{MFIM}
\bsm{I}_m(\bsm{\theta}')\triangleq\Expected\{\bsm{I}(\bsm{\theta}')\}=\begin{bmatrix}
\ \frac{2A^2NP_1}{\sigma^2\left(A^2|a|+1\right)}+\frac{A^4\Re\{a\}^2N}{|a|^2(A^2|a|+1)^2}& \frac{A^4\Re\{a\}\Im\{a\}N}{|a|^2(A^2|a|+1)^2}&0\\
\ \frac{A^4\Re\{a\}\Im\{a\}N}{|a|^2(A^2|a|+1)^2}& \frac{2A^2NP_1}{\sigma^2\left(A^2|a|+1\right)}+\frac{A^4\Im\{a\}^2N}{|a|^2(A^2|a|+1)^2} &0\\
\ 0&0&\frac{2A^2NP_2}{\sigma^2\left(A^2|a|+1\right)}
\end{bmatrix}.
\end{equation}
\hrulefill
\vspace*{3pt}
\setcounter{equation}{20}

\end{figure*}

\section{Simulation Results}
\label{simulations}

In this section, we compare the MSE performance of the ML estimator and the MSEV estimator using Monte-Carlo simulations. Our results are generated assuming $M=4$ (QPSK) and $P_r=P_1=P_2$, and are averaged over the same set of 100 realizations of the channel parameters $g$, $h$ which are independently generated from the complex Gaussian distribution with mean zero and variance 1. The minimizers of the objective functions in~\eqref{a_ML} and~\eqref{a_MSEV} are obtained using the steepest descent algorithm. The initial points are chosen using the sample average estimator, while the step size is chosen using backtracking line search~\cite{boyd2004convex}. As a reference, we also show the MSE performance for the two estimators when the solutions for~\eqref{a_ML} and~\eqref{a_MSEV} are obtained using grid search with a step size of $10^{-3}$. We also show the bounds $CRB_a$ and $MCRB_a$, where $CRB_a$ is averaged over many realizations of $\bsm{t}_1$ and $\bsm{t}_2$.

\begin{figure}[htbp]
\centering
\includegraphics[width=3.7in, height=2.8in]{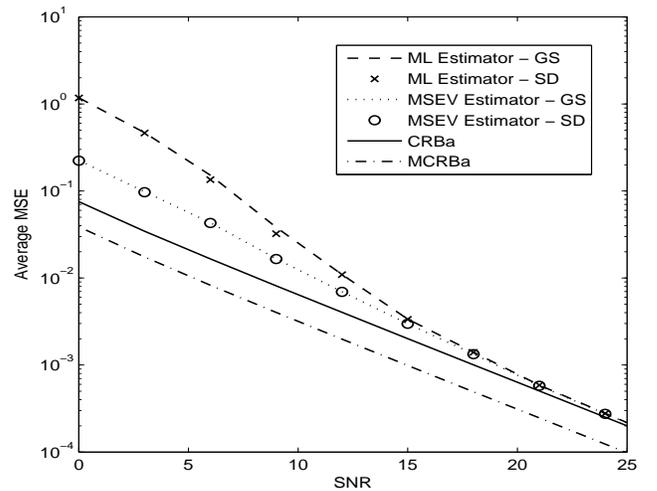}
\caption{Average MSE of the ML estimator and the MSEV estimator and the bounds $CRB_a$ and $MCRB_a$ vs transmit SNR for $N=100$, and $M=4$ (QPSK). For each estimator, two MSE plots are shown: one for the grid search (GS)-based channel estimate and one for the steepest descent (SD)-based estimate.}
\label{MSE_SNR}
\end{figure}

Fig.~\ref{MSE_SNR} shows the average MSE of the two estimators vs. SNR for $N=100$. The MSEV estimator outperforms the ML estimator and the performance gap is most significant at low to medium SNR. At high SNR, both estimators approach the bound $CRB_a$. We can also see from Fig.~\ref{MSE_SNR} that, for both algorithms, the MSE performance that results when steepest descent is employed is almost identical to that when grid search is used.

\begin{figure}[htbp]
\centering
\includegraphics[width=3.8in, height=3.8in]{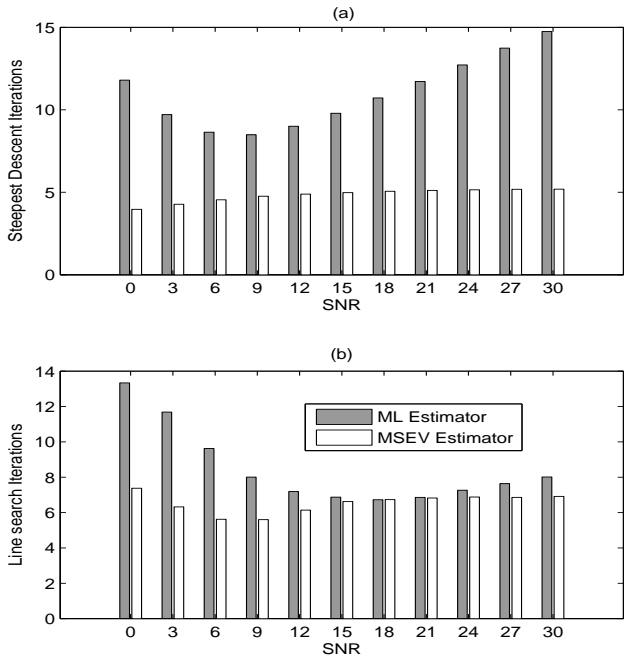}
\caption{Average number of steepest descent iterations (a), and average number of line search iterations for a single steepest descent iteration (b) required by the ML and MSEV estimators for different SNR values ($N=100$, $M=4$).}
\label{SD_LS_iterations}
\end{figure}

The bar plots in Fig.~\ref{SD_LS_iterations} show for both estimators the average number of steepest descent iterations required for convergence and the average number of line search iterations to find the step size for a single steepest descent iteration, respectively. As we can see from Figs.~\ref{MSE_SNR},~\ref{SD_LS_iterations}, the steepest descent algorithm is a reliable and low complexity method for solving~\eqref{a_ML} and~\eqref{a_MSEV}. Moreover, the MSEV estimator is more efficient than the ML estimator since on average it requires fewer steepest descent iterations to converge and fewer line search iterations to obtain the step size.

\begin{figure}[htbp]
\centering
\includegraphics[width=3.7in, height=2.8in]{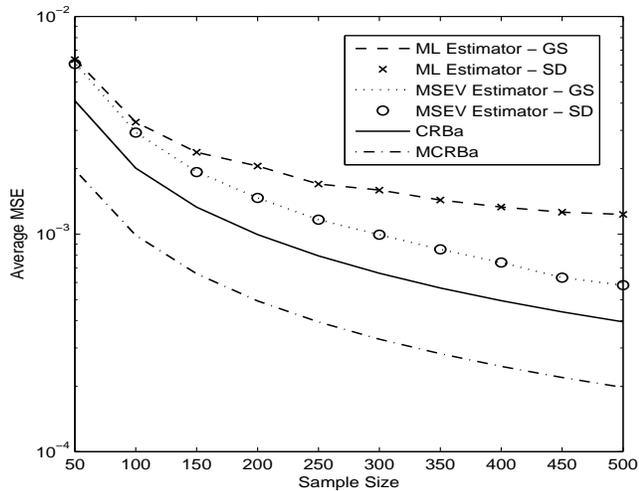}
\caption{Average MSE of the ML estimator and the MSEV estimator, and the bounds $CRB_a$ and $MCRB_a$ vs sample size for an SNR of 15 dB, and $M=4$ (QPSK). For each estimator, two MSE plots are shown: one for the grid search (GS)-based channel estimate and one for the steepest descent (SD)-based estimate.}
\label{MSE_N}
\end{figure}

Fig.~\ref{MSE_N} shows the average MSE of the two estimators vs. $N$ for an SNR of 15dB. The MSEV estimator has a superior MSE performance which improves steadily as $N$ increases. The gap between the MSE performances of the two estimators becomes more significant as the sample size increases, in accordance with the fact that the MSEV estimator is consistent while the ML estimator is not.

\section{Conclusions}
\label{conclusions}

In this work, we compared two partially-blind channel estimation algorithms for two-way relay networks assuming channel reciprocity and $M$-PSK modulation. The first was the ML estimator obtained by treating the data symbols as deterministic unknowns~\cite{ciss_paper}, and the second was the MSEV estimator which minimized the sample variance of the envelope of the received signal after self-interference cancellation. We showed that both estimators approach the true channel with high probability at asymptotically high SNR and that the MSEV estimator is consistent while the ML estimator is not. We also derived two CRBs on the variance of unbiased estimators. Monte-Carlo simulations were used to compare the MSE performance of the two estimators, showing that the MSEV estimator performs better than the ML estimator and that the steepest descent algorithm can be used to provide accurate low-complexity implementations for both estimators.

\appendices
\section{}
\label{appendix_ciss}
 We have
\begin{equation}
\label{app_ciss2}
\begin{split}
\hspace{-1.8ex}\left[1-\frac{\mathcal{W}(a)}{\sigma^2(A^2|a|+1)}\right]&\hspace{-0.5ex}=\frac{\pi}{4}\left(L_{1/2}\left(-\frac{A^2|b|^2P_2}{\sigma^2(A^2|a|+1)}\right)\right)^2\\
&\ \ -\frac{A^2|b|^2P_2}{\sigma^2(A^2|a|+1)}. 
\end{split}
\end{equation}
Let $\mathcal{Q}(x)\triangleq\frac{\pi}{4}\left(L_{1/2}\left(-x\right)\right)^2-x$, it is sufficient to show that $\mathcal{Q}(x)>0$ for $x>0$. In [9, Appendix E], it is shown that $\mathcal{Q}(x)$ is strictly decreasing for $x>0$. Using this fact, we will establish that $\mathcal{Q}(x)>0$ for $x>0$ by showing that $\lim\limits_{x\rightarrow\infty}\mathcal{Q}(x)=\frac{1}{2}$. We can expand $Q(x)$ as
\begin{equation}
\label{app_ciss3}
\begin{split}
Q(x)&=\frac{\pi}{4}e^{-x}\bigg[(1+x)^2I_0\left(\frac{x}{2}\right)^2+2x(1+x)I_0\left(\frac{x}{2}\right)I_1\left(\frac{x}{2}\right)\\
&\ \ \ \ \ \ \ \ \ \ \ \ \ \ \ \ \ \ \ +x^2I_1\left(\frac{x}{2}\right)^2\bigg]-x.
\end{split}
\end{equation}
Moreover, for large arguments, we have the following asymptotic expansion~\cite{abramowitz}
\begin{equation}
\label{appE10}
I_{\tau}(x)\approx\frac{e^{x}}{\sqrt{2\pi x}}\left\{1-\frac{\varepsilon-1}{8x}+\frac{(\varepsilon-1)(\varepsilon-9)}{2!(8x)^2}-\hdots\right\},
\end{equation}
where $\varepsilon=4\tau^2$. Using the above expansion, we obtain the following approximations for large $x$
\begin{equation}
\frac{\pi}{4}e^{-x}(1+x)^2I_0\left(\frac{x}{2}\right)^2\approx \frac{x}{4}+\frac{5}{8},
\end{equation}
\begin{equation}
\frac{\pi}{4}e^{-x}2x(1+x)I_0\left(\frac{x}{2}\right)I_1\left(\frac{x}{2}\right)\approx \frac{x}{2}+\frac{1}{4},
\end{equation}
\begin{equation}
\frac{\pi}{4}e^{-x}x^2I_1\left(\frac{x}{2}\right)^2\approx\frac{x}{4}-\frac{3}{8}.
\end{equation}
Substituting these approximations into~\eqref{app_ciss3}, we have $\lim\limits_{x\rightarrow\infty}Q(x)=\frac{1}{2}$, which completes the proof.

\bibliographystyle{IEEEtran}
\bibliography{IEEEabrv,twrn_bib}

\end{document}